\documentclass{epl}
\title{\Large{\bf{Signatures of two-dimensionalisation of 3D turbulence in presence of rotation}}}
\author{Sagar Chakraborty}
\institute{S.N. Bose National Centre for Basic Sciences, Theoretical physics department, Sector-III, Block - JD, Salt Lake, Kolkata - 700 098, India.}
\date{}
\usepackage{graphicx}
\usepackage{dcolumn}
\usepackage{bm}
\newcommand{\be}{\begin{eqnarray}}
\newcommand{\en}{\end{eqnarray}}
\newcommand{\ben}{\begin{eqnarray*}}
\newcommand{\enn}{\end{eqnarray*}}
\newcommand{\pa}{\partial}
\newcommand{\na}{\nabla}
\newcommand{\f}{\frac}
\newcommand{\td}{\tilde}

\newcommand{\p}{\paragraph{}}
\newcommand{\bi}{\begin{itemize}}
\newcommand{\ei}{\end{itemize}}

\newcommand{\la}{\langle}
\newcommand{\ra}{\rangle}

\renewcommand{\O}{\Omega}
\renewcommand{\p}{\bot}
\begin{document}
\maketitle
\begin{abstract}
A reason has been given for the inverse energy cascade in the two-dimensionalised rapidly rotating 3D incompressible turbulence.
For such system, literature shows a possibility of the exponent of wavenumber in the energy spectrum's relation to lie between -2 and -3.
We argue the existence of a more strict range of -2 to -7/3 for the exponent in the case of rapidly rotating turbulence which is in accordance with the recent experiments.
Also, a derivation for the two point third order structure function has been provided helping one to argue that even with slow rotation one gets, though dominated, a spectrum with the exponent -2.87, thereby hinting at the initiation of the two-dimensionalisation effect with rotation.
\end{abstract}
\section{Introduction}
Rotating turbulence, which is bridging the gap between 2D, quasi-2D and 3D turbulences, shows an extremely interesting property of two-dimensionalisation of 3D turbulence.
This aspect of research in turbulence is of current interest to oceanographers, geophysicists, meteorologists, mathematicians, physicists and others.
\\
\indent In the steady non-turbulent flow, for low Rossby number ($Ro=U/2L\O$) and high Reynolds number ($Re=UL/\nu$), Taylor-Proudman theorem\cite{Batchelor} argues that rotation two-dimensionalises the flow.
This argument is often mistakenly extended to turbulent flows to explain the rotation induced two-dimensionalisation arising therein.
The two-dimensionalisation of the 3D turbulent flow in presence of rotation has begun to be understood as a subtle non-linear effect, which is distinctly different from Taylor-Proudman effect, due to the works of Cambon\cite{Cambon1}, Waleffe\cite{Waleffe} and others who basically showed that all one means by the two-dimensionalisation is that the strong angular dependence of this effect leads to a draining of the spectral energy from the parallel to the normal wave vectors (w.r.t. the rotation axis).
Simulations ({\it e.g.} by Smith {\it et al.}\cite{Smith} which however deals with a particular case of forcing from the small scales) speak volumes for the two-dimensionalisation effect besides showing the initiation of inverse cascade of energy with rapid rotation, a fact well supported by the experiments\cite{Baroud1,Morize1}.
\\
\indent This reverse cascade of energy is an important signature of the two-dimensionalisation of turbulence due to rapid rotation and the other signature which we shall elaborately deal with in this letter is that the exponent of wavenumber in the energy spectrum's relation is different from the usual -5/3 for the isotropic homogeneous 3D incompressible turbulence.
\\
\indent Although recent experiments by Baroud {\it et al.}\cite{Baroud1,Baroud2} and Morize {\it et al.}\cite{Morize1,Morize2}  have shed some light on the two-dimensionalisation effect, the scaling of two-point statistics and energy spectrum in rotating turbulence remains a controversial topic.
An energy spectrum $E(k)\sim k^{-2}$ has been proposed\cite{Zhou,Canuto} for rapidly rotating 3D turbulent fluid and this does seem to be validated by some experiments\cite{Baroud1,Baroud2} and numerical simulations\cite{Yeung,Hattori,Reshetnyak,Muller}.
But some experiments\cite{Morize1} do not tally with this proposed spectrum.
They predict steeper than $k^{-2}$ spectrum and this again seem to be drawing some support from numerical results\cite{Yang,Bellet} and analytical results found using wave turbulence theory\cite{Galtier,Cambon2}.
\\
\indent However, one can always question the effectiveness of the signatures to be discussed because $i$) a scaling law for a single-component spectrum, though heavily used in literature, has poor meaning in the strongly anisotropic configuration relevant to pass from 3D-2D; different power laws can be found in terms of $k_z$, $k_{\bot}$ and $k$ in contrast to the 3D isotropic case, and $ii$) the inertial wave-turbulence theory is not consistent with an inverse cascade.
\section{The Signatures}
Let us first concentrate on why at all there should be an inverse cascade of energy.
Inverse cascade of energy is a trademark of 2D turbulence where a second conserved quantity -- enstrophy -- besides energy plays the defining role behind it.
One might be tempted to search for this conserved quantity in the case of rapidly rotating 3D turbulence, for, there in the limit of infinite rotation the axes of all the vortices are expected to point up towards the direction of angular velocity and looking at the every section perpendicular to the axis one might tend to take that as if showing 2D turbulence which obviously is not a correct inference because of the non-zero value of the axial velocity that may depend on the coordinates on the plane.
Searching for the enstrophy conservation seems to be a dead end as far as explaining the inverse cascade in rapidly rotating turbulence is concerned.
In such an unfortunate scenario, helicity (defined as $\int{\vec{v}}.{\vec{\omega}}d^3{\vec{r}}$) which remains conserved in a 3D inviscid unforced flow comes to our rescue.
It has been long known that helicity is introduced into a rotating turbulent flow\cite{Brissaud}.
Kraichnan\cite{Kraichnan} argued that both the helicity and energy cascade in 3D turbulence would proceed from lower to higher wave numbers and went on to remark that forward helicity cascade would pose a hindrance for the energy cascade, a fact validated by numerical simulations\cite{Andre,Polifke}.
He also showed that in presence of helicity two-way cascade is possible.
Lets see topologically why this should be so.
It is well known that a knotted vortex tube is capable of introducing helicity in fluid\cite{Moffatt}.
Consider a knotted vortex tube (see Fig-1) in a turbulent flow.
%
\begin{figure}
\centering
\includegraphics[width=7.0cm]{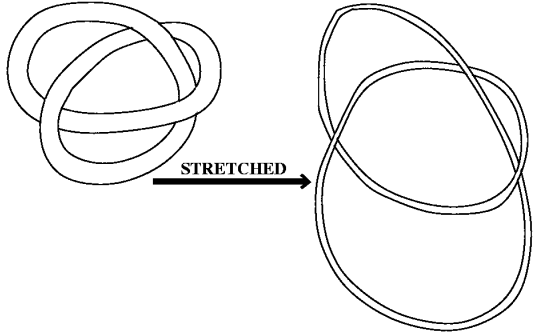}
\caption{A knotted vortex tube. When it is stretched the tube thins out to create smaller eddies but the entire structure occupies a larger volume.}
\end{figure}
%
Due to the vortex stretching phenomenon in turbulence, the vortex line stretches and as a result owing to the assumed incompressibility of the fluid the tube thins out keeping the volume inside it preserved and smaller scales are created; in a sense, this is what is meant by the flow of energy to the smaller scales.
But now this also means that the ``scale'' of the knotted structure would in general increase {\it i.e.,} the knot would now reach out to farther regions in the fluid.
Evidently, if we wanted this scale to reduce, we must let the stretched knotted tube fold in such a way so that the scale becomes smaller; such a neat arrangement seems to be a far cry in a turbulent flow which is inherently chaotically random causing the separation of two nearby particles of fluid on an average.
Thus, as the degree of knottedness measures helicity, the aforementioned argument suggests that if one forces energy to go to smaller scales, helicity would tend to go to larger scale and vice-versa.
This topological argument gives an intuitive way of comprehending how the forward helicity cascade can inhibit the forward cascade of energy.
The point is that in presence of forward helicity cascade, reverse cascade of energy is not impossible.
\\
\indent Waleffe\cite{waleffe2}, with the help of detailed helicity conservation by each triad, showed that helicity indeed affects the turbulence dynamics even in isotropic turbulence; this is a kind of catalytic effect.
One can thus take inspiration to make the argument in the previous paragraph more concrete by playing around with a simplified triad using logic in the line suggested by Fjortoft's theorem\cite{Fjortoft} in 2D turbulence.
Let the helicity spectrum be $H(k)$ and the energy spectrum be $E(k)$.
It may be shown that 
\be
|H(k)|\le kE(k)
\label{condition}
\en
Consider 3D Euler equation in Fourier space truncated in order to retain only three parallel wave vectors $\vec{k}_1$, $\vec{k}_2$ and $\vec{k}_3$ and suppose it is possible for these three particular wave vectors to be such that $|H(k)|=nkE(k)$, where $n$ is a positive number lesser than 1 to be in consistence with the relation (\ref{condition}).
Assume $\vec{k}_2=2\vec{k}_1$ and $\vec{k}_3=3\vec{k}_1$.
Conservation of energy and helicity imply that between two times $t_1$ and $t_2$, the variation $\delta E_i=E(k_i,t_2)-E(k_i,t_1)$ satisfies two constraints
\be
\delta E_1+\delta E_2+\delta E_3=0
\label{energy1}\\
nk_1\delta E_1+nk_2\delta E_2+nk_3\delta E_3=0
\label{helicity1}
\en
solving which in terms of $\delta E_2$, we get:
\be
\delta E_1=\delta E_3=-\f{\delta E_2}{2}
\label{energy2}\\
nk_1\delta E_1=-\f{n}{4}k_2\delta E_2;\phantom{xxx}nk_3\delta E_3=-\f{3n}{4}k_2\delta E_2
\label{helicity2}
\en
If one assumes that the wave vector $k_2$ is losing energy {\it e.g. $\delta E_2<0$}, then the results (\ref{energy2}) and (\ref{helicity2}) show that as more helicity goes into the higher wavenumber, the energy is equally transferred to both the lower and the higher wave numbers suggesting a possibility of the coexistence of reverse and forward energy cascades. 
\\
\indent Now let us come to the point.
In the case of 3D isotropic and homogeneous turbulence rotation can input helicity in it when there is a mean flow in the inertial frame and this value of input helicity increases with the increase in angular velocity.
Experiments on rotating turbulence invariably introduce helicity.
As the angular velocity is increased the helicity increases enough to inhibit the energy cascade appreciably so that a reverse cascade is seen.
This consistently explains the reason behind the existence of the reverse energy cascade in a rapidly rotating turbulent flow.
Hence, the argued existence of a direct helicity cascade in such experiments turns out to be an interesting (however not rigourously proven) assumption.
\\
\indent The next important signature of the two-dimensionalisation of turbulence that remains to be pondered over is the exponent of the wave vector in the energy spectrum relation.
To be precise, if one wishes angular velocity to become a relevant parameter in constructing the energy spectrum $E(k)$, simple dimensional analysis would lead one to:
\be
E(k)\propto\O^{\frac{3m-5}{2}}\varepsilon^{\f{3-m}{2}}k^{-m}
\label{0}
\en
where $m$ is a real number.
$m$ should be restricted within the range 5/3 to 3 to keep the exponents of $\O$ and $\varepsilon$ (rate of dissipation of energy per unit mass) in relation (\ref{0}) positive.
The two limits $m=5/3$ and $m=3$ corresponds to isotropic homogeneous 3D turbulence and 2D turbulence respectively.
The spectrum due to Zhou -- $E(k)\sim k^{-2}$ -- is due to an intermediate value of $m=2$.
So, as far as the present state of the literature on rotating turbulence goes, two-dimensionalisation of 3D turbulence would mean the dominance of a spectrum which goes towards $E(k)\sim k^{-3}$ and which may choose to settle at $E(k)\sim k^{-2}$, an issue yet to be fully resolved.
\\
\indent Lets give a twist to the tale.
In general, the energy spectrum\cite{Brissaud} in the inertial range will be determined by both the helicity cascade and the energy cascade which simply means that the energy spectrum from the dimensional arguments should be written as
\be
E(k)\propto\varepsilon^{\f{7}{3}-m}h^{m-\f{5}{3}}k^{-m}
\label{spectrum}
\en
where $h$ is the rate of helicity dissipation per unit mass.
Demanding positivity of the exponents of $\varepsilon$ and $h$, one fixes the possible values for $k$ within the closed range [5/3,7/3], imposing which on the arguments given in the previous paragraph, one can easily propound the range
\be
2\le m\le\f{7}{3}
\label{range}
\en
for the rapidly rotating 3D turbulent flow.
Direct experiments\cite{Morize1} by Morize {\it et al.} have found energy spectrum for rapidly rotating turbulence going as $k^{-2.2}$ which is as predicted by the relation (\ref{range}).
\section{Small Rotation Limit}
But one question still remains.
All the discussion mainly focused on turbulent flows with low $Ro$ and high $Re$.
What happens if the turbulent fluid is not rapidly rotated but is slowly rotated?
So, now to answer this question, lets focus on the slowly rotated isotropic homogeneous incompressible 3D turbulent fluid.
By slow rotation we mean that $\O\ll\sqrt{\varepsilon/\nu}$, $\nu$ being the kinematic viscosity.
It can be argued\cite{Sagar} that the coefficients of the general tensorial form for $b_{ij,k}\equiv\la v_iv_jv'_k\ra$ (where angular brackets mean ensemble average and $v_i=v_i(\vec{x},t)$ is the $i$-th component of velocity and similarly, $v'_i=v_i(\vec{x}+\vec{l},t)$) in the inertial range should depend explicitly on $l_z,\phantom{x}l_{\p}$ and $\O$.
So we have
\be
b_{ij,k}&=&C(l_\p,l_z,\O)\delta_{ij}l^o_k+D(l_\p,l_z,\O)(\delta_{ik}l^o_j+\delta_{jk}l^o_i)+F(l_\p,l_z,\O)l^o_il^o_jl^o_k
\label{bijk}
\en
where $l_i^o$ is the $i$-th component of the unit vector along $\vec{l}$.
It may be noted that, for simplicity, antisymmetric terms have been dropped.
Later on, in the end of this section we shall figure out the harm this simplification might have done.
One has the incompressibility condition:
\be
\pa'_kb_{ij,k}=0
\label{incompressibility}
\en
Using relation (\ref{incompressibility}) and relation (\ref{bijk}), we arrive at following relationships among the coefficients:
\be
D=-\f{l_\p}{2}\td{C}-\f{l_z}{2}\dot{C}-C\\
F=\f{l^2}{2}\td{\td{C}}+\f{l^2l_z}{2l_\p}\dot{\td{C}}+\left(\f{3l^2}{2l_\p}-\f{l_\p}{2}\right)\td{C}-\f{l_z}{2}\dot{C}-C
\label{F}
\en
Here tilde and dot define derivatives w.r.t. $l_\p$ and $l_z$ respectively.
So, one has
\be
B_{ijk}&\equiv&\la(v'_i-v_i)(v'_j-v_j)(v'_k-v_k)\ra\nonumber\\
&=&-2(l_\p\td{C}+l_z\dot{C}+C)(\delta_{ij}l^o_k+\delta_{ik}l^o_j+\delta_{jk}l^o_i)+6Fl^o_il^o_jl^o_k
\label{Bijk}
\en
And hence the two point third order structure function is
\be
S_3=B_{ijk}l^o_il^o_jl^o_k=6[F-(l_\p\td{C}+l_z\dot{C}+C)]
\label{S3}
\en
One may define physical space energy flux $\varepsilon(\vec{l})$ as:
\be
&&\varepsilon(l)\equiv-\f{1}{4}\vec{\nabla}_l.\la|\delta\vec{v}(\vec{l})|^2\delta\vec{v}(\vec{l})\ra
\label{eflux}
\en
And the energy flux $\Pi_K$ through the wave number $K$ for the homogeneous (not necessarily isotropic) turbulence may be shown to be\cite{Frisch}:
\be
\Pi_K=\f{1}{2\pi^2}\int_{{R}^3}d^3l\f{\sin(Kl)}{l}\vec{\nabla}_l.\left[\varepsilon(\vec{l})\f{\vec{l}}{l^2}\right]
\label{flux}
\en
Now if one probes into the small $l$ behaviour in the plane perpendicular to the rotation axis by putting $l_z=0$ and uses the relations (\ref{F}) to (\ref{flux}), doing tedious algebra one may land up on (details given elsewhere\cite{Sagar}):
\be
S_3|_{l_z=0}=-\f{6}{\pi}\varepsilon l_\p+Al_\p^{\f{7+\sqrt{97}}{6}}
\label{S3lz}
\en
where, $A$ is a constant which for obvious reason depends on $\O$ and $\varepsilon$.
Using dimensional arguments and introducing a non-dimensional constant $c$, we may set
\be
A=c\O^{\f{1+\sqrt{97}}{4}}\varepsilon^{\f{11-\sqrt{97}}{12}}
\label{A}
\en
From relations (\ref{S3lz}) and (\ref{A}), we may write finally
\be
S_3|_{l_z=0}=-\f{6}{\pi}\varepsilon l_\p+c\O^{\f{1+\sqrt{97}}{4}}\varepsilon^{\f{11-\sqrt{97}}{12}}l_\p^{\f{7+\sqrt{97}}{6}}
\label{finalS3}
\en
This (relation (\ref{finalS3})) is the final form for two-point third order structure function in the plane whose normal is parallel to the rotation axis for rapidly rotating homogeneous 3D turbulence.
\\
\indent Lets pause for a moment and summarise the assumptions and the steps involved in getting the relation (\ref{finalS3}) starting from the definition (\ref{bijk}) for the sake of completeness.
The rotation is taken to be low enough to ensure that $B_{ijk}$ in (\ref{Bijk}) could be written in terms of $b_{ij,k}$ and then the mild anisotropy introduced due to rotation is taken care of by writing $\la|\delta\vec{v}(\vec{l})|^2\delta\vec{v}(\vec{l})\ra=B_{ii\alpha}l^o_{\alpha}\vec{l}_{\bot}/{l}_{\bot}+B_{iiz}l^o_{z}\vec{l}_{z}/{l}_{z}$ (where $\alpha$ can take only two values -- $x$ and $y$).
This is coupled with (\ref{Bijk}) to use in the relation (\ref{eflux}) which in turn when input in (\ref{flux}) as an explicit function of $l_{\bot}$ and $l_z$ yields in the limit $\nu\rightarrow 0$ for the inertial scales a partial differential equation, the form of which in the limit of $l_{z}=0$ is
\be
\left[l_\p\f{\pa}{\pa l_\p}+1\right](3l_\p^2\td{\td{\td{C}}}+5l_\p\td{\td{C}}-12\td{C}-4\f{C}{l_\p})=-\f{8\varepsilon}{\pi}
\label{diffeqn}
\en
It is this equation which is solved to arrive at the relation (\ref{finalS3}) where $\varepsilon\equiv\lim_{\nu\rightarrow 0}\Pi_K$.
\\
\indent One can argue dimensionally that the relation (\ref{finalS3}) tells that in the directions perpendicular to the axis of rotation, there are two possible energy spectrums {\it viz.}
\be
&&E(k)\sim k^{-\f{5}{3}}
\label{26}\\
\textrm{and}, \phantom{xxx}&&E(k)\sim k^{-\f{16+\sqrt{97}}{9}}
\label{27}
\en
which are respectively due to the first term and the second term in the R.H.S. of the relation (\ref{finalS3}).
Obviously, the spectrum (\ref{26}) will be dominant compared to the spectrum (\ref{27}).
But as the $\O$ is increased, the spectrum (\ref{27}) becomes more and more prominent; thereby two-dimensionalisation of the 3D homogeneous turbulent fluid is initiated which then carries over to high rotation regime.
It is very interesting to note that the exponent of $k$ in the relation (\ref{27}), {\it i.e.} $-(16+\sqrt{97})/9$, equals $-2.87$ which is in between $-3$ (for 2D turbulence) and $-2$ (for rapidly rotating 3D turbulence as proposed by Zhou).
It hasn't fallen into the more strict range [-7/3,-5/3] obviously because $\O$ is too low and may be because to maintain isotropy to a certain extent for the sake of hiccup-free calculations we have chosen not to include terms involving $\epsilon_{ijk}$ in the relation (\ref{bijk}) which could grab the effect of helicity explicitly; thereby again showcasing the need for the helicity to be effective to give the right exponent for the rotating turbulence.
\section{Yet Another Signature}
 Having explained the two signatures of the two-dimensionalisation effect, we search for another possible signature of the effect.
The advection of a passive scalar $\theta$ may serve the purpose since the Yaglom's law\cite{Monin} in d-D incompressible turbulent fluid may be written as $\la\delta v_{\parallel}(\delta\theta)^2\ra=-({4}/{d})\varepsilon_{\theta}l$, where $\varepsilon_{\theta}\equiv\kappa\left\la\left({\pa_{l_i}}\theta\right)\left({\pa_{l_i}}\theta\right)\right\ra=-{\pa_t}\la\theta^2\ra$ and $\kappa$ being the diffusivity.
This law distinguishes between a 2D and a 3D turbulence and hence it is worth getting a form for it for a rotating 3D turbulence and find if in a plane perpendicular to the rotation axis it reduces to the form for 2D turbulence and thereby bringing in the effect of two-dimensionalisation.
Since one can show that small $\O$ could bring in anisotropy in the otherwise isotropic scales\cite{Sagar}, one would look out for the effect of small $\O$ on the passive scalar which follows the equation:
\be
\f{\pa\theta}{\pa t}+\vec{\na}.(\vec{v}\theta)=\kappa\na^2\theta-\epsilon_{ijk}\O_j\f{\pa}{\pa x_i}(x_k\theta)
\label{ps}
\en
If one goes by the proof given in the reference \cite{Biskamp} to find out a value for $\la\delta v_{\parallel}(\delta\theta)^2\ra$ for small $l$ in this case assuming very small $\O$ (and hence isotropy), one arrives back at the Yaglom's law.
We can however land up on a very neat experimentally and numerically verifiable correlation which can serve the purpose of a signature of two-dimensionalisation if we treat equation (\ref{ps}) anisotropically as follows.
\\
\indent Defining $\vec{l}\equiv\vec{x'}-\vec{x}$ and $\pa_{l_i}\equiv\na_i=\pa'_i=-\pa_i$, one can manipulate the equation (\ref{ps}) to get:
\be
\pa_t\la(\delta\theta)^2\ra+\na_i\la\delta v_i(\delta\theta)^2\ra=2\kappa\na_{ii}\la\theta^2\ra-4\kappa\la\na_i\theta\na_i\theta\ra-\epsilon_{ijk}\O_j\na_i\la l_k(\delta\theta)^2\ra
\label{psr}
\en
Now, owing to the anisotropy caused by rapid rotation, we may write $\la\delta\vec{v}(\delta\theta)^2\ra=\la\delta{v_{\bot}}(\delta\theta)^2\ra\vec{l}_{\bot}/{{l}_{\bot}}+\la\delta{v_{z}}(\delta\theta)^2\ra\vec{l}_{z}/{{l}_{z}}$ and as $\la(\delta\theta)^2\ra$ is proportional to terms quadratic in $l_{\bot}$ and $l_z$, in the limit $\kappa\rightarrow0$ and small scales, one can easily reach at the following relation:
\be
\la\delta v_{\bot}(\delta\theta)^2\ra|_{l_{\bot}=0}=0
\label{psc}
\en
This relation predicts that in the presence of rapid rotation, and hence anisotropy, on the small line segment parallel to axis of rotation the correlation in the L.H.S. of (\ref{psc}) vanishes. This may be readily used in numerics to check if the two-dimensionalisation has been achieved and hence may be treated as a signature of the effect.
\section{Conclusion}
To conclude, we mention that true reason behind the so called two-dimensionalisation of turbulence has been figured out which accounts for the energy cascade direction and the energy spectrum found in the experiments and simulations.
To settle the problem more neatly, study of passive scalars in rotating turbulence may prove to be of benefit which of course is our future course of action.
\acknowledgements
Prof. J.K. Bhattacharjee is gratefully acknowledged for the many fruitful and rewarding discussions. Also, CSIR (India) is acknowledged for awarding fellowship to the author. The anonymous referees are thanked for their positive criticisms.

\end{document}